\documentclass[aps,pre,longbibliography,twocolumn,floatfix]{revtex4-1}
\usepackage{graphicx}
\usepackage{amsmath}
\usepackage{amssymb}
\usepackage{bm}
\usepackage{times}
\usepackage[pdftex]{hyperref}
\hypersetup{colorlinks=true,linkcolor=blue,citecolor=blue,urlcolor=blue}
\usepackage{verbatim}

\begin{document}

\title{The realizable solutions of the TAP equations}

\author{T.~Aspelmeier}
\affiliation{Institute for Theoretical Physics,  Georg-August-Universit\"at G\"ottingen,  D37077, G\"ottingen,  Germany }

\author{M.~A.~Moore}
\affiliation{Department  of
Physics and Astronomy, University of Manchester, Manchester M13 9PL, UK}
\date{\today}

\begin{abstract}
We show  that the only  solutions of the  TAP equations for  the
Sherrington-Kirkpatrick model  of Ising spin glasses  which can be  found by
iteration  are those  whose free  energy  lies on  the border  between
replica symmetric and broken replica symmetric states, when the number
of  spins $N$  is large.  Convergence to  this same  borderline also
happens in quenches from a high temperature initial state to a locally
stable state where each spin is  parallel to its local field; both are
examples of self-organized criticality. At this borderline the band of
eigenvalues of the Hessian associated with a solution extends to zero,
so  the  states   reached  have  marginal  stability.   We  have  also
investigated the  factors which  determine the free  energy difference
between a stationary solution corresponding  to a saddle point and its
associated minimum, which is the barrier which has to be surmounted to
escape from  the vicinity of a  TAP minimum or pure state.

\end{abstract}
\maketitle

\section{Introduction}
\label{sec:introduction}
One of the most influential papers in the theory of spin glasses was the paper of Thouless, Anderson and Palmer (TAP) \cite{thouless:77}. They provided a set of $N$ coupled equations for the magnetization $m_i$ at site $i$ of the Sherrington-Kirkpatrick (SK) \cite{sherrington:75} model of Ising spin glasses. Since then equations equivalent to those of TAP have been studied for  $p$-spin models, which are models for structural glasses, and also for a host of computer science applications \cite{mezard:17}.

The free energy $F$ (multiplied by $ \beta= 1/k_BT$) associated with a TAP state for the Ising spin SK model is
\begin{eqnarray}
F&=&-\beta \sum_{i <j} J_{ij} m_i m_j -\frac{\beta^2 N}{4}(1-q)^2 \nonumber\\
&+& \sum_i \left[\frac{1+m_i}{2} \ln \frac{1+m_i}{2}+\frac{1-m_i}{2}\ln \frac{1-m_i}{2}\right],
\label{TAPF}
\end{eqnarray}
where $q =(1/N) \sum_i m_i^2$. The TAP equations themselves are derived from the stationarity equations $\partial F/\partial m_i=0, i=1,\ldots,N$ and take the form
\begin{equation}
m_i =\mathcal{G}_i({m})=\tanh[\beta \sum_{j \ne i}  J_{ij} m_j-\beta^2(1-q)m_i].
\label{TAP}
\end{equation}
The Hessian associated with the stationary points of $F$ was studied long ago \cite{Bray:78a, aspelmeier:06}. It is defined by
\begin{eqnarray}
&A_{ij}&=\frac{\partial^2 F}{\partial m_i \partial m_j}\nonumber\\
&=&-\beta J_{ij}- \frac{2 \beta^2}{N}m_i^* m_j^* +\big[\frac{1}{1-(m_i^*)^2}+\beta^2(1-q)\big]\delta_{ij}, \nonumber\\
\label{Hessian}
\end{eqnarray}
where $m_i^*$ denotes the magnetization at site $i$ at a stationary point.

A great deal is already  known about the solutions of the TAP equations and their associated Hessians. There are an exponentially large number of solutions for $\beta > 1$, that is $T < T_c=1$. The complexity of the minima of $F$ is defined by
\begin{equation}
\Sigma_{\rm min}(f) =\frac{\ln N_{\rm solns}(f)}{N},
\label{complexitydef}
\end{equation}
where $N_{\rm solns}(f)$ denotes the number of minima of  free energy per spin $f=F/\beta N$.  $\Sigma_{\rm min}(f)$ quantifies the number of solutions when it is exponentially large. It is non-zero over a range of $f$ values  \cite{bray:80,Bray:81, aspelmeier:04,cavagna:04}. The solution of lowest free energy per spin $f_0$ is one of the pure states of the Parisi replica symmetry broken solution (RSB) \cite{parisi:79, mezard:87}. The pure states are those whose free energies per spin are only $O(1/N)$ above $f_0$. There is a critical $f_c$ at which the solutions change their nature \cite{bray:80,Bray:81} . Those solutions at $f > f_c$ are uncorrelated with each other and their Hessians have a single, nearly null eigenvalue, (whose value vanishes in the limit $N \to \infty$), separated  by a gap from a band of $N-1$ eigenvalues. As the free energy is reduced towards $f_c$ the gap goes to zero and for all $f\le f_c$ there is no gap between the lowest ``null'' eigenvalue and the bottom of the band \cite{aspelmeier:04}. The states with $f < f_c$ have non-trivial RSB overlaps with each other \cite{bray:80,Bray:81}. Those with $f> f_c$ have trivial (zero) replica symmetric (RS) overlap with each other.

A stationary solution of the TAP equations which corresponds to a minimum will have all the eigenvalues of its associated Hessian non-negative. It turns out that the other stationary points are saddle points with just  a single negative eigenvalue \cite{aspelmeier:04}. Every minimum has its associated saddle point, so the complexity of the saddle points is identical to that of the minima. Once over the saddle-point in the direction away from the minimum one goes  towards the trivial paramagnetic solution ($P$) for which  $m_i^*=0$. Thus the free energy landscape of the free energy functional of Eq.~(\ref{TAPF}) is simple; it has a spoke-like arrangement of minimum, associated saddle and $P$ at the center of the wheel  \cite{aspelmeier:04,cavagna:04}. The free energy difference between the free energy at the saddle point and its associated minimum is the barrier which has to be surmounted to escape from the minimum. In this paper we have once more  investigated these barriers in order to determine some of the factors which might control their magnitude. The simulations of Billoire et al.  \cite{billoire:01,Billoire:10} indicate that the barriers separating pure states may scale as $N^{1/3}$. Our studies suggest that non-pure state solutions will have much smaller barriers, of $O(1)$, so that it will be possible to escape from their vicinity by thermal fluctuations (in agreement with our earlier study  \cite{aspelmeier:06}).

The chief purpose of this paper is to report a feature of the actual solutions found in numerical work which has not previously been noticed. In numerical work, the free energy minima can be obtained by the iterative map:
\begin{equation}
m_i^{(k+1)}=m_i^{(k)}+ \alpha[\mathcal{G}_i(m^{(k)})-m_i^{(k)}],
\label{iteration}
\end{equation}
where $\alpha$ is a parameter which controls the approach to the next iterate \cite{aspelmeier:06}. In this work, we used $\alpha = 1.2$ except where indicated. The value of $\alpha$ affects  the free energy per spin  $f$ which is obtained, and instead of obtaining a spread of values of $f$ the iteration scheme picks out in the large $N$ limit one particular value of $f$, $\tilde{f}$. What we want to point out is that our states at this particular value of $f$ have the properties of states at the critical value $f_c$, which is the borderline between states which are replica symmetric (RS) and those whose replica symmetry is broken (RSB) \cite{bray:80,Bray:81}. $\tilde{f}$ is less than $f_c$.  $f_c$ is that associated with all possible minima of the TAP functional, rather than the subset produced by the chosen iteration scheme.  In our work the states found lie close to $\tilde{f}$, differing from it by an amount which decreases as $N$ becomes large. Even though $\tilde{f}$ corresponds to a free energy per spin below $f_c$, (so it nominally lies in the region where the states would have RSB features), the states produced in the iteration (which are just a subset of all the possible states at $\tilde{f}$) do not have this feature. Instead the subset of states generated is closer in its properties to those at the borderline $f_c$ itself.

 The iteration procedure of Eq.~(\ref{iteration}) is just one of a large number of ways of solving the TAP equations, but we suspect that any iterative solution will have the same features as those found using Eq.~(\ref{iteration}). Evidence for this belief is to be found in Ref. \cite{plefka:03}.  Plefka used an iterative scheme which was similar to solving  Glauber dynamical equations. He found that the states which he obtained all had the same free energy and which were also marginal, just as we find. (His scheme converges to a value of the free energy different  to the one which we find). We have also studied the iterative procedure of Bolthausen \cite{Bolthausen:14}. In his procedure the Onsager reaction term was calculated at stage $k-1$.  Bolthausen was able to show that in a field the procedure converged in the paramagnetic phase but we found it less efficient  than that of Eq.~(\ref{iteration}) in converging to a stationary solution: starting from some initial state a common feature is just bouncing around without convergence. However, Eq.~(\ref{iteration}) was more likely to find a  solution than the Bolthausen method at large $N$ values and we have used it throughout this paper.

What led us to carry out this investigation was the work of Sharma et al. \cite{sharma:18,sharma:16}. It was found in these papers that a quench from an initial random (i.e. high-temperature) spin configuration by successively putting spins in turn parallel to their local fields until all are so aligned led to a final quenched state which in SK type models lay on the boundary between replica symmetric states and states with RSB. This is what we also find for solutions of the TAP equations; the iterative solution has parallels with the quenching procedure. The number of  quenched states at $T=0$ has also  been studied as a function of their energy, and there exists a critical energy per spin $e_c$ below which the states have RSB features and above which the states are uncorrelated \cite{bray:80,Bray:81b}. A problem with studying the Ising model at $T=0$, i.e. in the quenched state, is that a Hessian cannot be constructed as the spins take the values $\pm 1$, so marginality as indicated by eigenvalues of a Hessian extending down to zero \cite{muller:15}, cannot be investigated. A big disadvantage of studying the finite temperature TAP equations is that the values of $N$ which can be studied with the TAP equations are much smaller than those which can be handled in a quench. The same self-organized critical features are present in both the solutions of the TAP equations and in the quenched states  and presumably the physics behind this is the same \cite{sharma:18}, i.e.\ somewhat obscure, at least to us. However, for certain aspects of the quenched problem one has some features which are rigorously established; Newman and Stein \cite{newman:99} have shown that in the large $N$ limit, the quench takes one to a particular energy per spin $\tilde{e}$ which is self-averaging, but dependent on the algorithm used. It would be nice if their proof could be extended to the somewhat similar TAP problem, as our work shows that as $N$ gets large that there is  convergence to a particular free energy $\tilde{f}$.

In a recent study Montanari \cite{montanari:19} has discussed an algorithm  which returns a spin configuration $\{S_i\}$, $S_i =\pm 1$, such that the energy $E=-\sum_{i<j} J_{ij} S_i S_j$ lies above the true ground state energy $E_g$  by an amount $-\epsilon E_g$.  It works by utilizing the full Parisi replica symmetry solution. It is our suspicion that using such $\{S_i\}$ as starting points,  TAP solutions quite different from those we study could be generated. Our initial state is a random state, similar to a paramagnetic state and our solutions have no overlap with each other. Our work has similarities to a quench to zero temperature from infinite temperature. We suspect though that Montanari's procedure extended to generate TAP solutions might be similar to a quench from an initial equilibrated state at $T < T_c$ where replica symmetry breaking features are present in the initial state and which are presumably retained during the quench. If TAP solutions can be generated from the initial state $\{S_i\}$ we suspect that they will have non-trivial overlaps with each other.

The details of our numerical work can be found in Sec.~\ref{sec:simulation} while in Sec.~\ref{sec:SOC} we present the evidence that in the large $N$ limit the TAP solutions which can be found lie at the boundary between solutions whose overlaps are replica symmetric and those whose overlaps are those of broken replica symmetry. Our work on barriers is in Sec.~\ref{sec:barriers}. We have fitted the free energy between the minimum and the saddle with a quartic fit and as a consequence can relate the barrier height to the difference in the values of $q$ at the minimum and the saddle, and the curvatures at the minimum and the saddle. We then discuss how the barriers between pure states could become of order $N^{1/3}$. Finally in Sec.~\ref{sec:discussion} we comment upon unresolved issues.  In Appendix A we have derived a finite size correction to the position of the Hessian band-edge, which turns out to work well at the rather modest values of $N$ which we can study.

\section{Simulation details}
\label{sec:simulation}
We studied the TAP equations for $N = 20, 40, 80, 160$ and $320$ spins and 500 bond realizations for each $N$. For each realization we tried to find solutions by iteration  according to Eq.~\eqref{iteration}, starting from a random initial state
\begin{align*}
  m_i &= \tanh(\beta \sqrt{q_s}X_i),
\end{align*}
where $q_s$ is the (fictitious) replica symmetric value of $q$ which is the solution of $q=\int \tanh(\beta\sqrt q \xi)^2\exp(-\xi^2/2)\,d\xi/\sqrt{2\pi}$ \cite{sherrington:75}, and where $X_i$ are normally distributed random variables. This construction ensures an initial state with a value of $q$ roughly in the range of typical TAP solutions.
As mentioned in Sec.~\ref{sec:introduction} we used  $\alpha=1.2$ except for the final approach (see below). We chose the temperature $T=0.3$  as a compromise between having too small a probability of finding any solution at all,  as happens for $T$ close to $T_c$, and having $q \approx 1$, which is the case for  $T$ close to 0. The latter would lead to  complications by causing a very large spread in the eigenvalues of the Hessian, as discussed in \cite{aspelmeier:06}, obfuscating the issues we are focusing on here.

In order to avoid questions of numerical accuracy, which can be very delicate in the  complex TAP free energy landscape, we used arbitrary precision arithmetic with 512 binary digits for the final approach to a TAP solution and for subsequent calculations.
The final approach is done in terms of the transformed variables $x_i = -\,\mathrm{sign}(m_i)\log(1-m_i^2)$ by iterating a transformed version of Eq.~\eqref{iteration} (see Eq.~(12) in \cite{aspelmeier:06}) and with $\alpha=1$ since the final approach starts off already in the basin of attraction. Use of the transformed variables is necessary because the original $m_i$ may take the values $\pm 1$ within numerical accuracy upon iterating Eq.~\eqref{iteration} directly, in which case the Hessian is ill defined, see Eq.~\eqref{Hessian}.

For each solution found we tried to locate the corresponding saddle using the method described in \cite{aspelmeier:04}. We then calculated  the eigenvalues and eigenvectors of the Hessian at both the minimum and the saddle (if it was found). Since the Hessian matrices can be very ill-conditioned,  it is in this step where the arbitrary precision arithmetic is most useful.

Since the quantities we examine  in this work may have strongly non-Gaussian distributions, such as for instance the low-lying eigenvalues of the Hessian, with possibly fat tails, we used the median instead of the mean consistently throughout this work for robustness. Accordingly, all error bars shown are 95\% confidence intervals for the median.

\section{Self-organized criticality}
\label{sec:SOC}
\begin{figure}
  \includegraphics[width=\columnwidth]{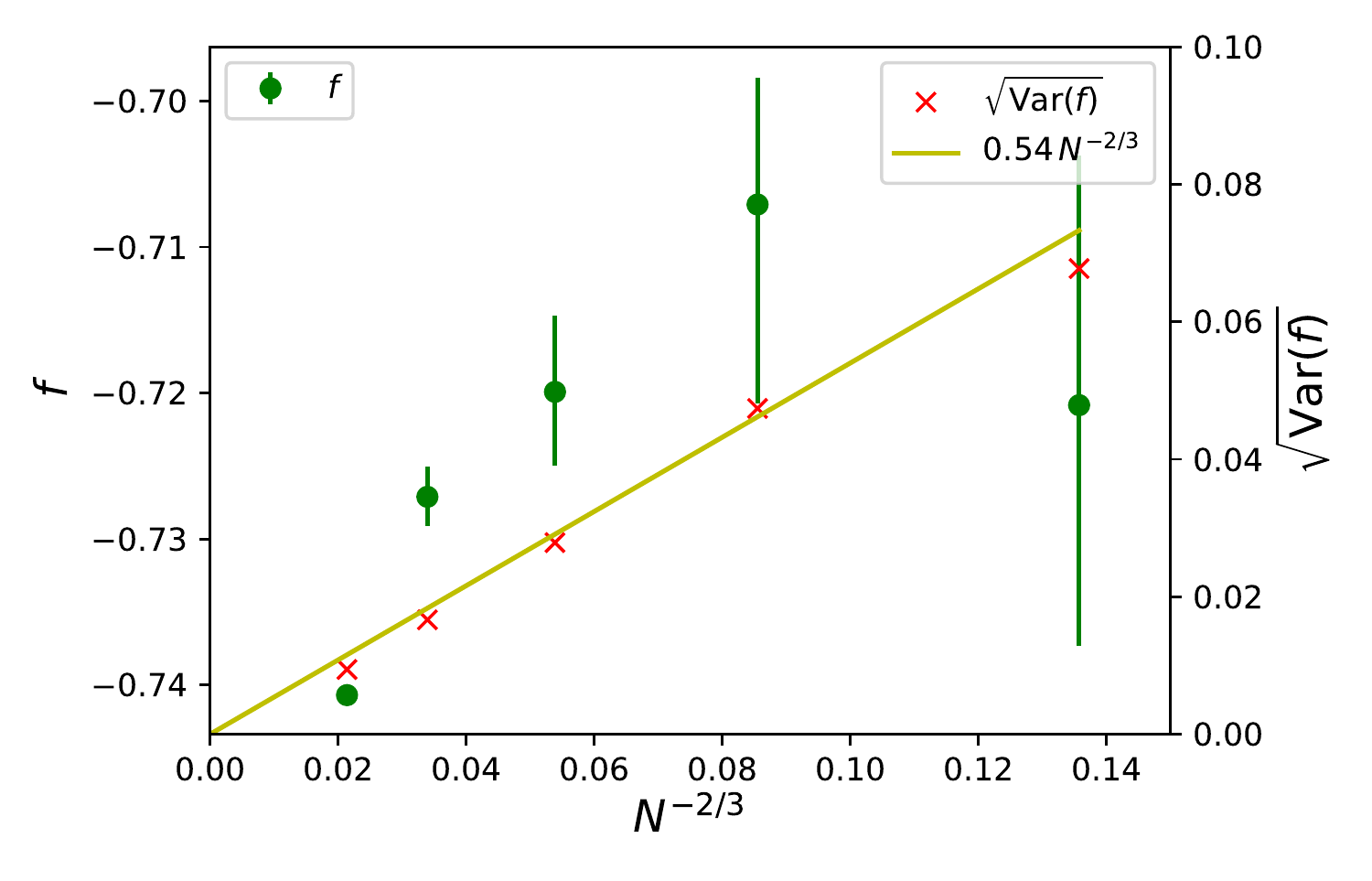}
  \caption{The free energy per spin, $f$, of the TAP solutions   and the square root of its variance  plotted against $1/N^{2/3}$ at a temperature $T=0.3$. }
  \label{fig:fvsN}
\end{figure}

In this section we give the details of why we believe  that the solutions of the TAP equations which are found by an iterative process lie at the boundary between replica symmetric and broken replica symmetry solutions. In Fig.~\ref{fig:fvsN} we have plotted the free energy per spin of the solutions found at a temperature $T=0.3$ and with $\alpha =1.2$ as a function of $1/N^{2/3}$. The variance decreases strikingly rapidly  as $N$ increases, suggesting that as $N \to \infty$, there will be a well-defined limit for the free energy $\tilde{f}$. In Ref.~\cite{cavagna:04} the entire $\Sigma_{{\rm min}}(f)$ curve was obtained when studying values of $N$ up to $80$. Unfortunately as $N$ increases this becomes harder and harder to do as the solutions found are approaching $\tilde{f}$ and solutions well away from this value are rarely found. Thus the authors of Ref.~\cite{cavagna:04} only succeeded in finding all the solutions by virtue of finite size effects. As $N$ grows, the chance of finding solutions well away from $\tilde{f}$  rapidly decreases to zero.  The rapid decrease of the variance with increasing $N$ is very suggestive that the solutions being found do not come from all over the $\Sigma_{{\rm min}}(f)$ curve, (which would lead to an $N$ independent variance) but rather are just those associated with a particular value of the free energy. We do not have any arguments as to why the $N$ dependence of $f$ and the square root of its  variance should vary as $1/N^{2/3}$;  we only use a $1/N^{2/3}$ as this form arises frequently for finite size effects in the SK model \cite{aspelmeier:06,aspelmeier:08}. The data for $f$ is not even monotonic  as a function of $1/N^{2/3}$ which suggests that the values of $N$ which we can study are not yet large enough to be in the asymptotic regime for this quantity.

\begin{figure}
  \includegraphics[width=\columnwidth]{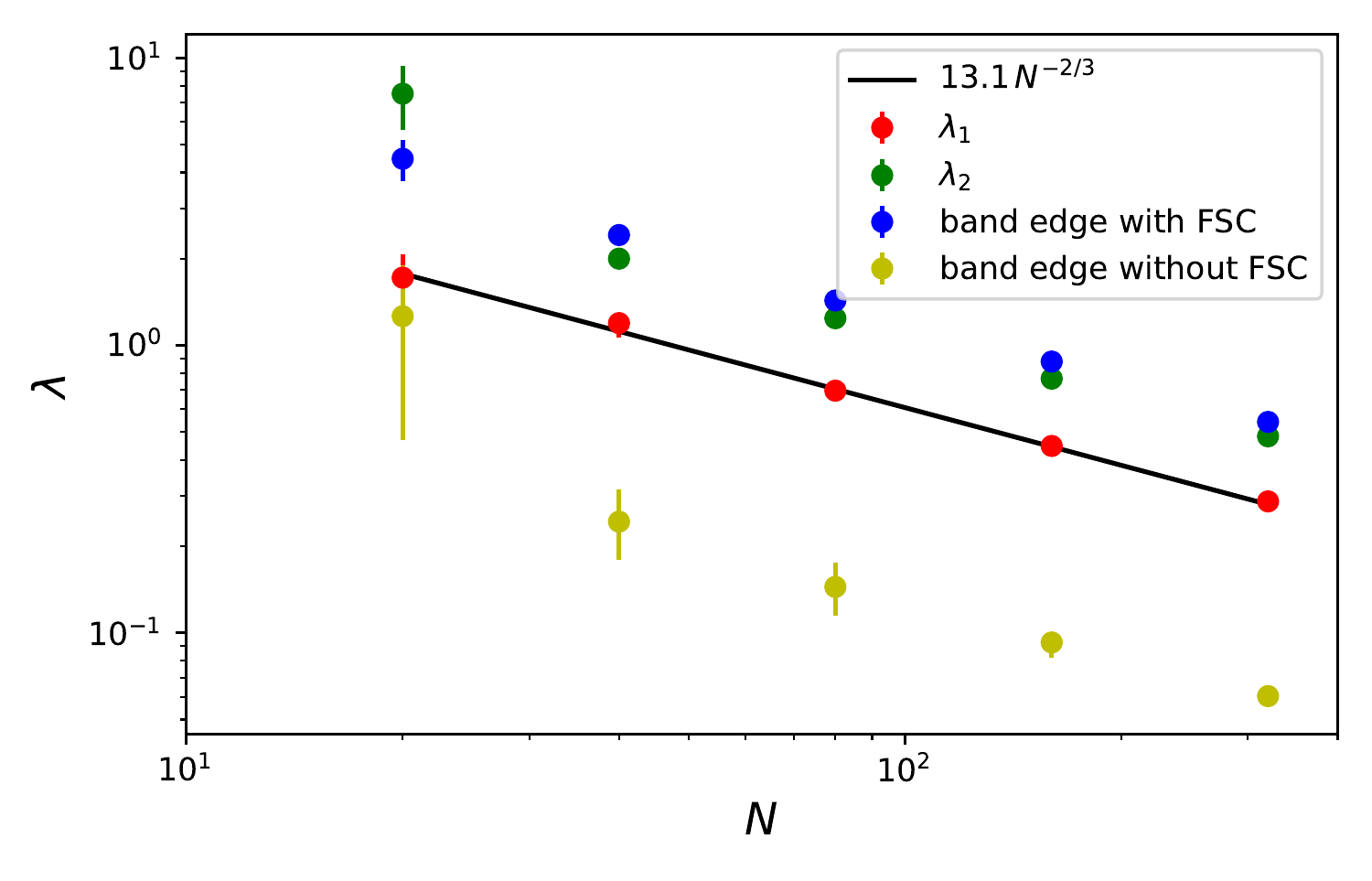}
  \caption{The lowest two eigenvalues $\lambda_1$ and $\lambda_2$ versus $N$ on a log-log scale. All data are for $T=0.3$ with $\alpha=1.2$. The black line shows a line of slope $1/N^{2/3}$.  Also shown are the results  for the band-edge $\lambda_2$ without finite size corrections (FSC) and the band-edge with finite size correction, both discussed in Appendix \ref{app:bandedge}; the latter is closer to the observed values of $\lambda_2$. }
  \label{fig:lambdaN}
\end{figure}

Our contention is not only that the solutions found in an iterative procedure converge to a unique value of the free energy as $N \to \infty$ but the particular free energy converged to is the \textit{critical} free energy which separates states with vanishing overlaps from those with non-trivial overlaps. We shall refer to this borderline as the RS/RSB critical point. The free energy per spin  $f_c$ is the free energy at this borderline when all possible minima of the TAP equations are studied. The subset of these states which we obtain by iteration whose free energies are close to  $\tilde{f}$ have the features of states at $f_c$. We have obtained the Hessians associated with the minima obtained by iteration.  Fig.~\ref{fig:lambdaN} shows that the two lowest eigenvalues of the Hessian seem to be both approaching zero as $1/N^{2/3}$. The smallest eigenvalue $\lambda_1$ is the ``null'' eigenvalue associated with the broken supersymmetry \cite{aspelmeier:04,cavagna:04,rizzo:04}. The second eigenvalue $\lambda_2$ lies at the bottom of the band of eigenvalues of the Hessian and for states with $f> f_c$ should be different from $\lambda_1$ by a finite amount which does not vanish as $N \to \infty$. Notice that for the smallest value of $N$ in the plot, $N=20$, where one will be sampling states over a wide range of $f$ values, one can see that indeed $\lambda_2$ looks quite distinct from $\lambda_1$.

 \begin{figure}
 \includegraphics[width=\columnwidth]{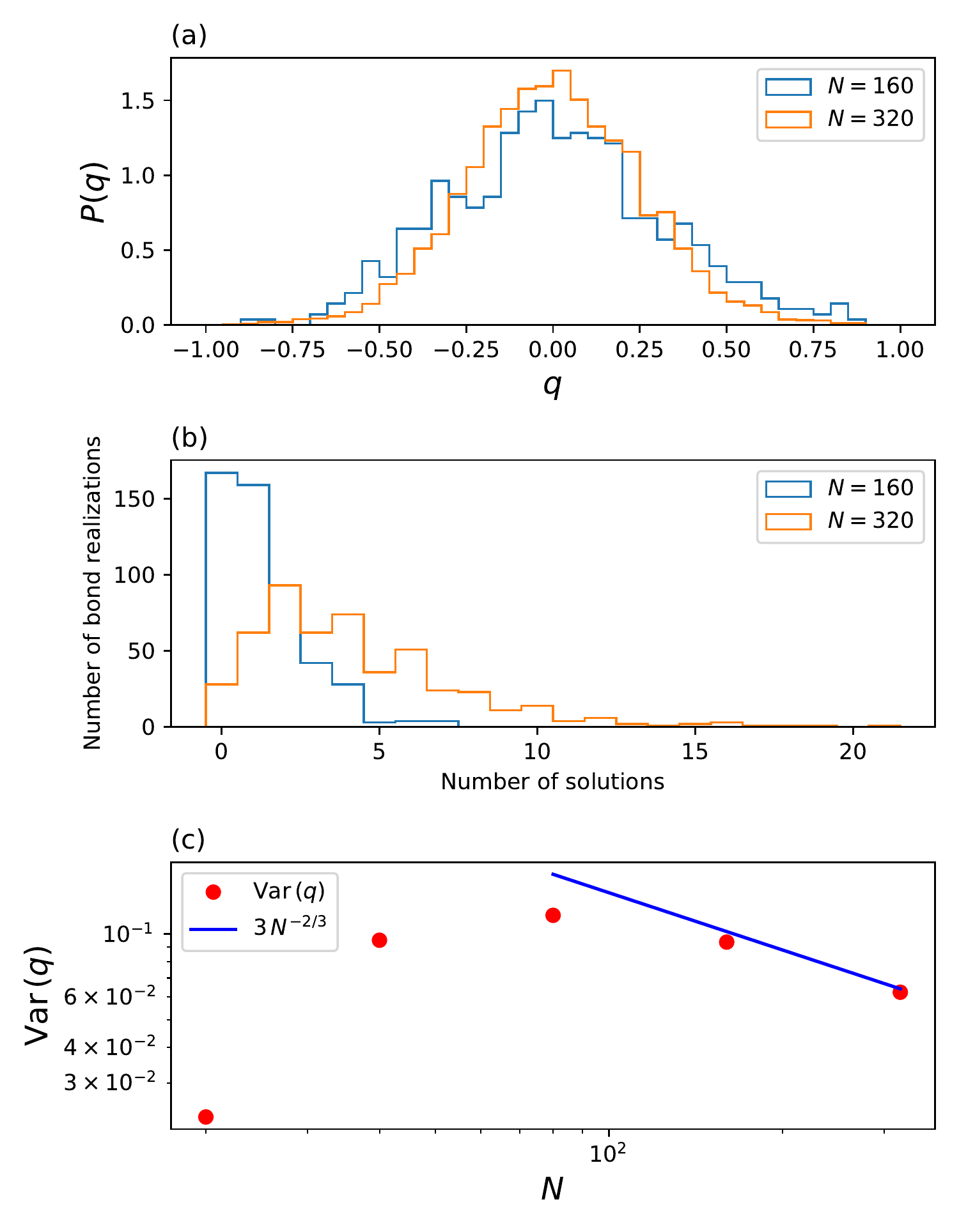}
  \caption{(a) Histograms for the probability density of overlaps of the solutions found at two different $N$ values, $N=160$ and $N=320$. The distribution of $q$ is expected to shrink towards $P(q) =\delta(q)$ as $N\to \infty$. The histograms are compiled from the overlaps of all pairs of solutions belonging to the same bond realization, averaged over all realizations (see  Eq.~(\ref{P(q)def})). All realizations with at least 2 solutions were used. (b) Histograms of the numbers of solutions.  (c) The variance of the overlaps $q$ plotted against $N$. For  $N> 80$ a shrinkage perhaps to zero is becoming visible.  }
 \label{fig:P(q)}
\end{figure}

The bottom of the band can be calculated by considering the matrix $X_{ij}$ defined via $A_{ij}=(X^{-1})_{ij}- (2 \beta^2/N) m_i m_j$, i.e.\ the Hessian without the projector term. The projector term, being $\mathcal O(1/N)$, is only a small perturbation which changes the eigenvalues of $A$ only slightly except for the isolated one. We expect that $\lambda_2\approx \mu_2$, the second smallest eigenvalue of $X^{-1}$.
If we define $p= \beta^2 N^{-1}\sum_i(1-m_i^2)^3$, then the band-edge without finite size corrections should be
at  \cite{Bray:78a, Plefka:02}
\begin{equation}
 \lambda_2=x^2/(4 p) \hspace{0.15cm} {\rm where} \hspace{0.15cm}
 x=1-\beta^2 N^{-1}\sum_i (1-m_i^2)^2.
\label{lambda2a}
\end{equation}
This is, however, only an approximation valid for small $x^2/(4p)$. In Appendix \ref{app:bandedge} we show how to find the exact individual band edge $z_0$ numerically.
We have plotted $z_0$ for $\lambda_2$ in Fig.~\ref{fig:lambdaN} but the agreement with the measued values of $\lambda_2$ is not good, presumably because of finite size effects.
In Appendix \ref{app:bandedge} we describe how to obtain a finite size correction for the band-edge, which does indeed improve the  agreement with $\lambda_2$.

The results in Fig.~\ref{fig:lambdaN} show that as $N$ increases both $\lambda_1$ and $\lambda_2$ are approaching zero, indicating that the solutions we are finding in this limit are similar to those whose  free energy is less than $f_c$. Below $f_c$ the states are associated with full replica symmetry breaking \cite{Bray:81} and would be associated with massless modes so that for all TAP states with   $f < f_c$ one would expect both  $\lambda_1$ and $\lambda_2$ to decrease as $1/N^{2/3}$, just as found in Fig.~\ref{fig:lambdaN}.
To see that the convergence is not to a state below $f_c$ but to a state right at the bordeline between RS and RSB states, we have studied the overlaps of the solutions in Fig.~\ref{fig:P(q)}.

The probability density of overlaps of  solutions at a given size $N$ was defined as
\begin{equation}
P(q)=\frac{1}{N(N-1)}\sum_{s \ne s'} \delta(q-N^{-1} \sum_i m_i^s m_i^{s'}).
\label{P(q)def}
\end{equation}
(Note that $q$ here is not that of the TAP equation;  $s$ and $s'$ denote two distinct solutions). We have also averaged the result over $J_{ij}$ realizations. Fig.~\ref{fig:P(q)}(a) shows $P(q)$ for two different system sizes. Both peak at $q=0$, a feature which would not be expected when $f < f_c$.  Right at $f=f_c$ the expected  form of $P(q)= \delta(q)$ in the large $N$ limit. We expect that this peak is broadened by finite size effects so that the data at finite $N$ and $q$ could be collapsed onto a universal curve by plotting against $q N^{1/3}$, but we do not have data at large enough values of $N$ to study this.  Fig.~\ref{fig:P(q)}(c) shows the variance of $q$ shrinks with $N$ for  $N> 80$, which is what would be expected if $P(q)$ is approaching a delta function at large $N$. For states with $f< f_c$ the variance of $q$ would be expected to approach a non-zero value in the large $N$ limit. Fig.~\ref{fig:P(q)}(b) shows the number of bond realizations for which a given number of solutions was found. For $N= 160$, the most common number of solutions found was zero!  For $N=320$ the situation  improves, presumably because the larger the value of $N$ the more solutions there are to be found. Fig.~\ref{fig:P(q)}(c) illustrates why the $N$ values which we can reach are a long way away from being in the large $N$ regime for some quantities, and Fig.~\ref{fig:P(q)}(b) illustrates how hard it is to get non-trivial solutions of the TAP equations.

The fact that $\lambda_2 \sim 1/N^{2/3}$  also explains another puzzling feature associated with solving the TAP equations by iteration; the solutions have always been reported from the earliest days as having a Hessian spectrum whose band-edge extended down to zero \cite{Bray:78a}, rather than having a finite band gap as expected, for example, at the peak of the $\Sigma_{\rm min}(f)$ or for any $f > f_c$.  In fact, no finite band gap has ever been clearly seen in numerical studies of the TAP equations.

We conclude that the solutions which are found by iteration are at an RS/RSB border. They are an example of self-organized criticality. The states (solutions) are associated with a Hessian whose eigenvalues extend to zero, and so are also \textit{marginal} \cite{muller:15} as well as self-organized.

\section{Barriers}
\label{sec:barriers}
In the SK model, the low-temperature spin glass phase has broken replica symmetry. That means it is associated with pure states, whose free energy per spin differ from each other by terms of $O(1/N)$ \cite{mezard:87}. Escape from a pure state is prevented  by large barriers. The simulations in Refs.~\cite{billoire:01,Billoire:10} indicate that the barriers  scale with the number of spins $N$ as $N^{1/3}$. Unfortunately there seems to be only a little understanding of the origin or form of these barriers \cite{Rodgers:89}. In this paper we shall try to cast some light on them by assuming that TAP solutions whose free energies per spin are within $O(1/N)$ of $f_0$, the solution of lowest free energy, can be identified as pure states and that the barrier for escaping a pure state can be identified with the free energy difference between the free energy of a TAP minimum and its associated saddle point.
Alas, as pointed out in Sec.~\ref{sec:SOC} the only states which we can find by directly solving the TAP equations are those which are around a free energy on the RS/RSB boundary (i.e. around $\tilde{f}$) and not those whose free energies lie within  $O(1/N)$ of $f_0$.  However, by examining the factors which  determine the magnitude of barriers we have been able to understand the features which  have to be present for barriers to scale as $N^{1/3}$.

\begin{figure}
  \includegraphics[width=\columnwidth]{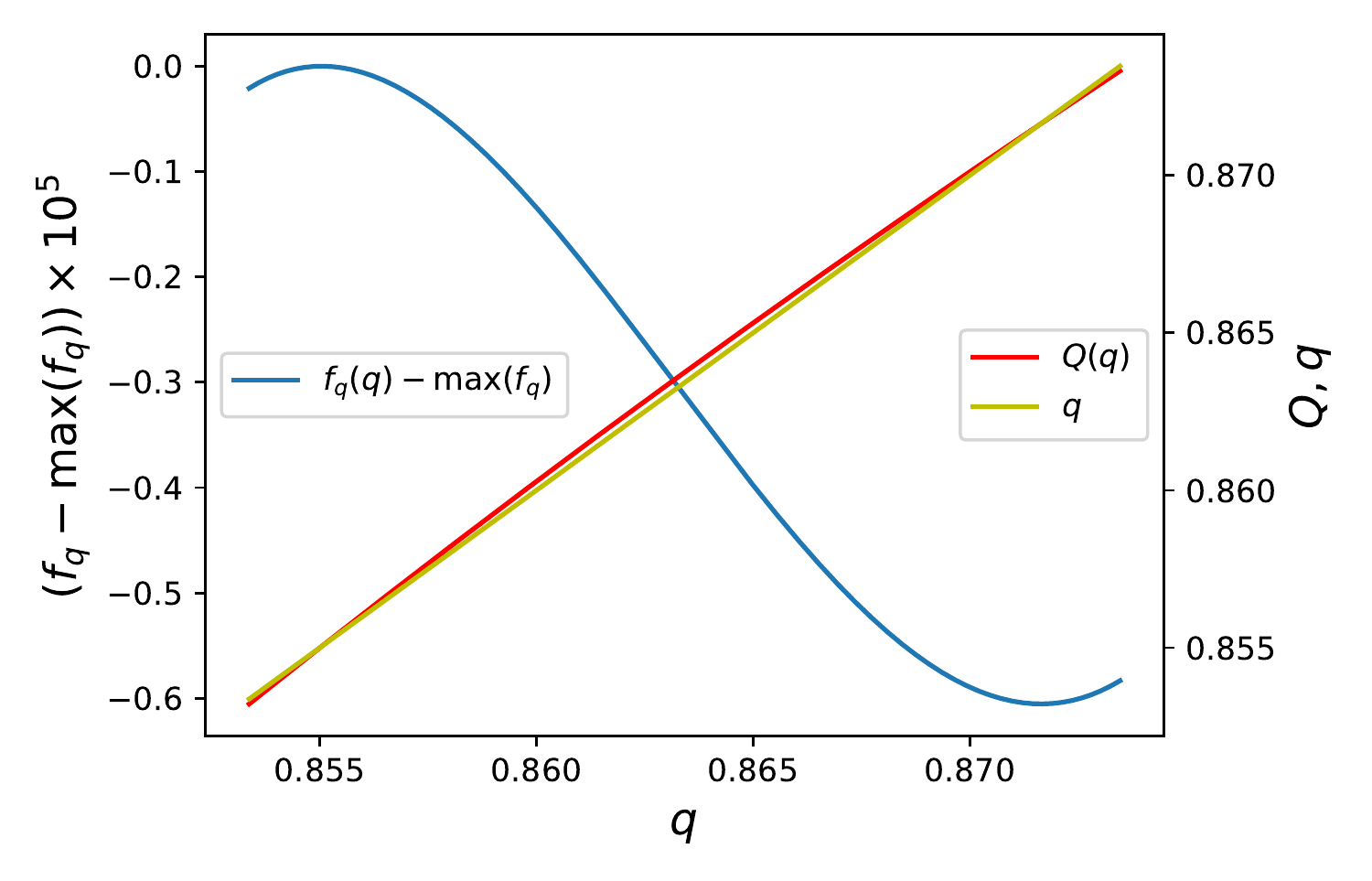}
  \caption{The functions $Q(q)$ (red line) and $f_q-\mathrm{max}(f_q)$ (blue line) versus $q$. The minimum and the saddle of $f_q$ occur where $Q(q)$ crosses the yellow line $Q=q$ for $N= 320$. The free energy per spin at the saddle is $\mathrm{max} (f_q)$.}
 \label{fig:f_q(q)}
\end{figure}

The TAP free energy $F_q$ as a function of $m_i$ and $q$ is defined \cite{aspelmeier:04} via
\begin{equation}
F_q=F+\frac{\beta^2}{2}(1-q)(\sum_im_i^2-Nq),
\label{def:Fq}
\end{equation}
where $F$ is the functional of $m_i$ and $q$ of Eq.~(\ref{TAPF}) except that now $q$ is regarded as an independent variable, unrelated to the $m_i$; $F_q$ is a function of the $N+1$ variables $m_1,\ldots,m_N,q$, whereas the original TAP free energy $F$ depends only on the variables $m_1,\ldots,m_N$ (with $q$ \textit{defined} as $q=(1/N) \sum_i m_i^2$).  One can easily show that the stationarity equations for $F_q$ reproduce the TAP equations: $\partial F_q/\partial m_i=G_i=0$. However for these new equations the quantity $Q\equiv\frac 1N\sum_i m_i^2$ is in general not equal to the parameter $q$ appearing in the equations. However, the additional stationarity equation, $0=\partial F_q/\partial q=(\beta^2/2)(N q-\sum_im_i^2)$ forces $Q=q$ at stationary points in the full $(N+1)$-dimensional space. Therefore at the minimum and the saddle the free-energy functions $F$ and $F_q$ have the same $m_i$ and $q$ values. By formally eliminating the variables $m_i$ by use of the TAP equations one can obtain the function $F_q$ as a function of $q$. Starting from a minimum, where $Q=q$, and following $F_q(q)$ for decreasing $q$ until $Q=q$ again, the corresponding saddle is found. In Fig.~\ref{fig:f_q(q)} we have plots of $f_q=F_q/N$ and $Q$ as functions of $q$.

To understand how the barrier height, which is the free energy difference between the saddle-point value $F_s$ of the free energy and the minimum value $F_m$, i.e.\ $B =F_s-F_m$, might depend on the values of $q_m-q_s$ and the curvatures at the minimum and the saddle, we have  used a quartic fit to $F_q$:
\begin{eqnarray}
 \tilde{F}&=& c \left[\frac{b}{4} (q-q_s)^4+\frac{1}{3}(q-q_s)^3-\frac{a}{2}  (q-q_s)^2 \right] \nonumber\\
 &\approx& F_q -F_s,
\label{def:quartic}
\end{eqnarray}
where we will relate the coefficients $c$, $b$ and $a$ to the curvatures at the saddle and the values of $q$ at the minimum $q_m$ and at the saddle, $q_s$.
$\tilde{F}$ is stationary when
\begin{equation}
\partial \tilde{F}/\partial q=c\left[b (q-q_s)^3+(q-q_s)^2-a (q-q_s)\right]=0.
\label{stationary}
\end{equation}
$q_m$ is the solution of $b (q_m-q_s)^2 +(q_m-q_s)- a=0$. The free energy $\tilde{F}_s$ at the saddle is zero and at the minimum
\begin{equation}
\tilde{F}_m=c (q_m-q_s)^2\left[ \frac{b }{4}(q_m-q_s)^2+\frac{1}{3}(q_m-q_s)-\frac{a}{ 2}\right].
\label{fmin}
\end{equation}
The barrier is
$B =\tilde{F}_s -\tilde{F}_m = -\tilde{F}_m$.
 The curvature at the saddle is defined as
\begin{equation}
\partial^2 \tilde{F}/\partial q^2 = c\left[3 b (q-q_s)^2+2 (q-q_s) -a\right],
\label{curvs}
\end{equation}
 and equals $-ac$ at the saddle $q=q_s$, while  at the minimum
\begin{equation}
\partial^2 \tilde{F}/\partial q^2=c\left[3 b (q_m-q_s)^2+ 2 (q_m-q_s) -a\right].
\label{curvm}
\end{equation}

\begin{figure}
  \includegraphics[width=\columnwidth]{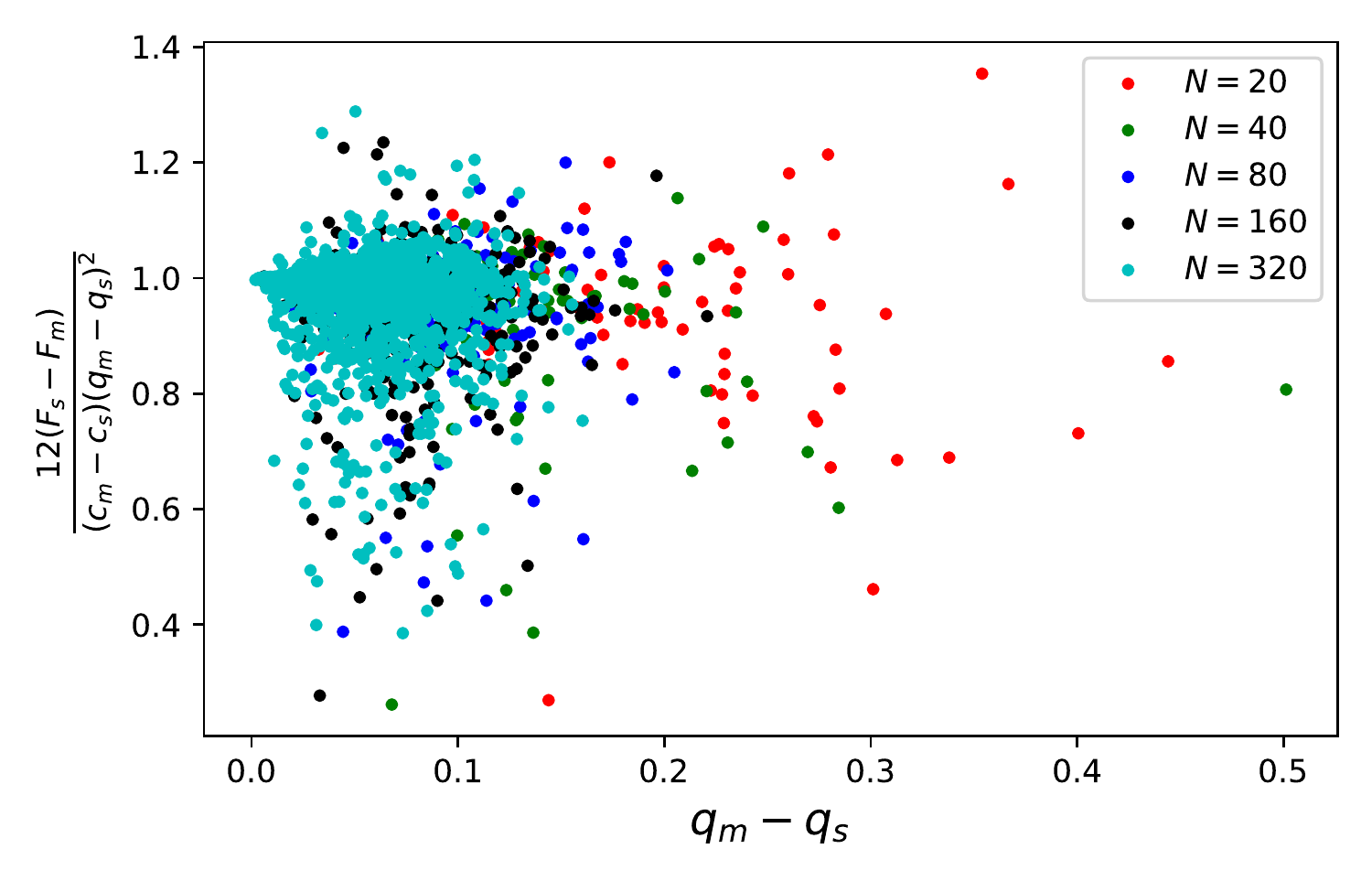}
  \caption{Plot of the barrier height $(F_s-F_m)$ divided by the right hand side of Eq.~(\ref{barrier}) versus $q_m-q_s$. The closeness to unity of this ratio indicates  the accuracy of the quartic fit  for a particular saddle-minimum pair.  }
  \label{fig:barrierplot}
\end{figure}

These curvatures at the saddle $-ac=c_s$ and at the minimum $c\left[3 b (q_m-q_s)^2+ 2 (q_m-q_s) -a\right] =c_m$, where the curvatures $c_s$ and $c_m$ were discussed in Ref.~\cite{aspelmeier:04};
\begin{equation}
 c_s \hspace{0.15cm} \mathrm{or} \hspace{0.15cm} c_m= \frac{N\beta^2}{2}\big(1-\frac{2 \beta^2}{N}\sum_{ij} m_i X_{ij}m_j\big),
\label{cs}
\end{equation}
evaluated for $X_{ij}$ at values of $m_i$ at the saddle or the minimum.

 We can eliminate the coefficients $a$, $b$, and $c$, to get
\begin{equation}
B=\frac{1}{12} (q_m-q_s)^2 (c_m-c_s).
\label{barrier}
\end{equation}
In Fig.~\ref{fig:barrierplot} we have plotted the observed barrier divided by the right hand side of Eq.~(\ref{barrier}) to check the accuracy of this equation. It clearly works well for most saddle-minima pairs, but a few are clearly not well-accounted for by the quartic fit of Eq.~(\ref{def:quartic}). For these pairs the discrepancy is simply because the neglected higher terms are just not always negligible.

\begin{figure}
  \includegraphics[width=\columnwidth]{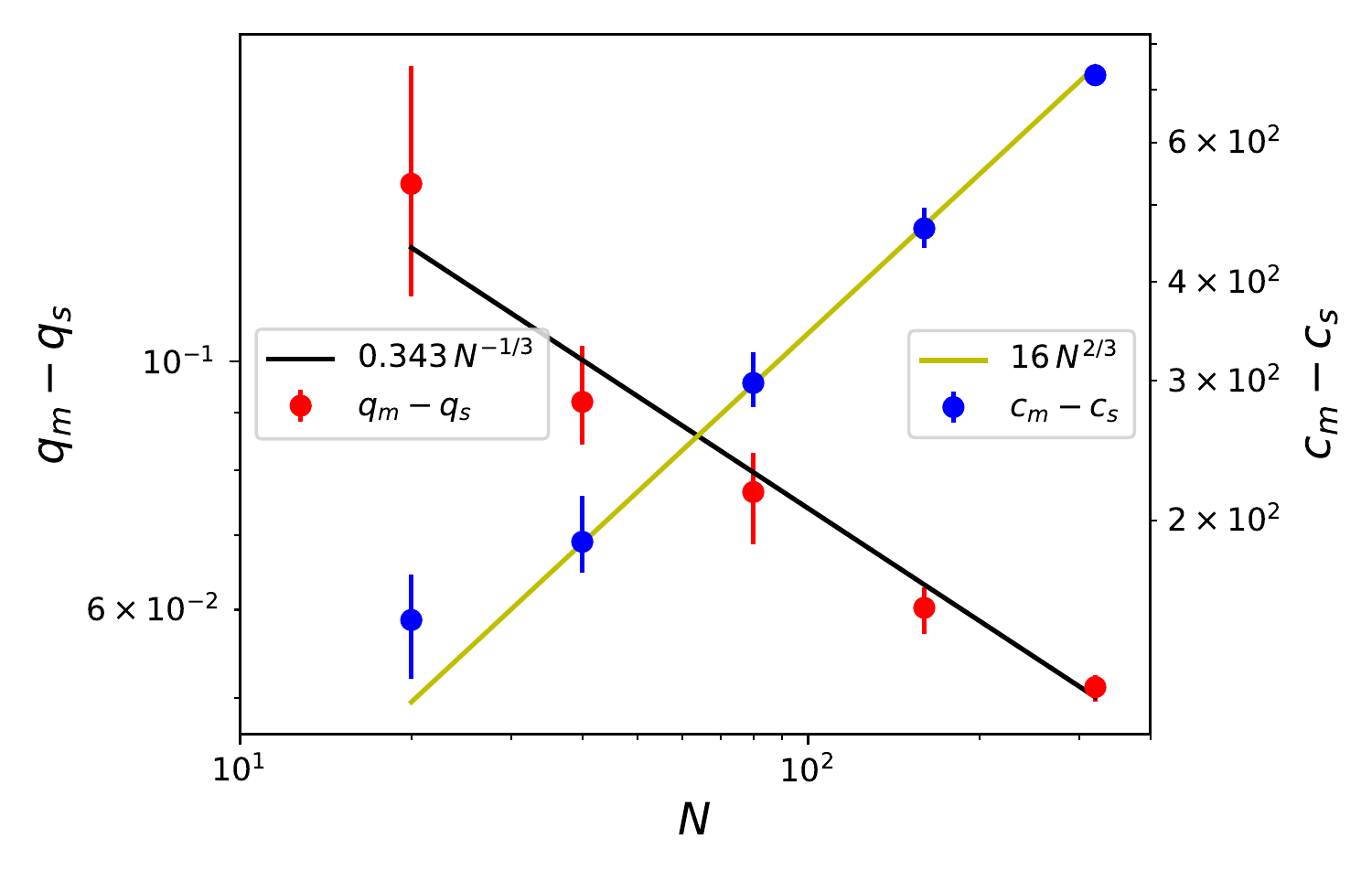}
  \caption{Plot of $q_m-q_s$ on the left axis and $c_m-c_s$ on the right axis, both on a logarithmic scale versus $N$, again on a log scale. Lines of slope $ \sim N^{-1/3}$ for $q_m - q_s$ and of slope $\sim N^{2/3}$ for $c_m-c_s$ have been drawn.  }
  \label{fig:Ndeps }
\end{figure}

Assuming that the quartic fit provides a good fit to the barrier height $B$ we next  describe the  $N$ dependence of the terms in Eq.~(\ref{barrier}).   In Fig.~\ref{fig:Ndeps } evidence is presented that $(q_m-q_s)$ decreases with $N$ as $\sim 1/N^{1/3}$ while the curvatures $(c_m -c_s)$ grow like $\sim N^{2/3}$. The variation of $(q_m-q_s) \sim 1/N^{1/3}$ means that the saddle becomes very close to the minimum in the large $N$ limit. Eq.~(\ref{barrier}) then implies  that the barrier height $B$ should be $N$ independent. In Ref. \cite{aspelmeier:06} we showed by varying the iteration parameter $\alpha$ that the barriers were $N$-independent, varying as  $ B \sim 1/(f-f_0)^{1/3}$. Hence at the critical free energy $\tilde{f}$ between RS/RSB states, the barriers would be expected to be $N$-independent. For pure states $f-f_0 \sim O(1/N)$, which explains why pure states have barriers of order $N^{1/3}$. Note that at $\tilde{f}$, the free energy associated with our iterative solutions, the barriers are numerically tiny, as can be seen from Fig.~\ref{fig:f_q(q)}.

\section{Discussion}
\label{sec:discussion}
While the SK model is referred  to as a ``solvable'' model, the finite
size corrections  to the thermodynamic  limit have only  been obtained
for a few quantities from analytical work. Mostly all that we have are
rather unsatisfactory  estimates from  numerical studies. The  same is
true  of the  TAP equations.  They become  exact in  the thermodynamic
limit, but finite $N$ corrections to  them and the $N$ dependencies in
their solutions have  not really emerged from  analytical studies. TAP
solutions of very  low free energies correspond to the  pure states of
the SK  model, in that if  one could compute  a value of $m_i$  in the
pure  state   it  would   correspond  to  that   of  a   TAP  solution
\cite{dominicis:83, bray:84}.

Our main discovery is that at large values of $N$ the solutions of the TAP equations fall at the boundary between states with replica symmetric overlaps and those with overlaps like those of broken replica symmetry. This is like a critical point. These states are associated with massless modes at large $N$ and so the solutions found are those of a self-organized marginally  stable critical system.

We do not know how  this behavior comes about.  But as the same behavior arises in quenched states of the SK model, it seems there exists a phenomenon worthy of further study.


\appendix

\section{Individual band edge and its finite size correction}
\label{app:bandedge}

In this Appendix, we derive the form of the finite size corrections to the individual band-edge which was used in constructing Fig.~\ref{fig:lambdaN}. By individual we mean  for a given TAP solution.

The eigenvalue density $\rho$ of $(X^{-1})_{ij} = -\beta J_{ij} +\big[\frac{1}{1-(m_i^*)^2}+\beta^2(1-q)\big]\delta_{ij}$, i.e.\ the Hessian without the projector term, can be obtained from its resolvent
\begin{align*}
  R(z) &= \frac 1N \,\mathrm{Tr}(z-X^{-1})^{-1}
\end{align*}
as
\begin{align}
  \rho(\mu) &= \frac{1}{\pi}\lim_{\epsilon\searrow 0}
  \,\mathrm{Im}\, R(\mu - i\epsilon).
  \label{eq:evdens}
\end{align}
As explained in \cite{Plefka:02}, the resolvent $R'$ of $X^{-1}/\beta$ satisfies the equation
\begin{align}
  R'(z) &= \frac 1N \sum_i \left(z - R'(z) - \beta^{-1}(1-m_i^2)^{-1}
  - \beta(1-q)\right)^{-1}
  \label{eq:pastur}
\end{align}
in the large $N$ limit according to Pastur's theorem \cite{Pastur:73}. The two resolvents are related by $\beta R(\beta z) = R'(z)$, hence $R$ satisfies
\begin{align}
  R(z) &= \frac 1N \sum_i \left(z - \beta^2 R(z) - (1-m_i^2)^{-1}
  - \beta^2(1-q)\right)^{-1}
  \label{eq:pastur2}
\end{align}
after a change of variables $\beta z \to z$.
Using a quadratic approximation to Eq.~\eqref{eq:pastur} valid for small $z- R'(z)-\beta(1-q)$, one obtains \cite{Plefka:02}
\begin{align*}
  \rho(\mu) &= \frac{1}{\pi \beta^{2}\sqrt p}\sqrt{\mu-x^2/4p}
\end{align*}
for small $\mu$ and $x^2/4p$, where $x$ and $p$ are defined as in Sec.~\ref{sec:SOC}. The band edge is thus at $x^2/4p$ \footnote{Note that it says $x^2/p$ in \cite{Plefka:02} for the band edge, which is a typo. Note also that our definition of the free energy, and hence also the band edge, differs from Plefka's by a factor of $\beta$ which is hidden in our definition of $p$.}.

However, $x^2/4p$ is not always small in our numerical experiments. Hence we refined this approximation by searching numerically for the infimum of real $z$ for which Eq.~\eqref{eq:pastur2} has no appropriate real solution, as this marks the onset of the band of eigenvalues according to Eq.~\eqref{eq:evdens}. To this end, define $Y \equiv \beta R(z)-z/\beta$ and $ k_i \equiv \beta^{-1}(1-m_i^2)^{-1}+\beta(1-q)$; Eq.~\eqref{eq:pastur2} then reads
\begin{align}
  z  &= -\beta\left(Y  + \frac 1N \sum_i \left(Y+k_i\right)^{-1}\right).
  \label{eq:resmax}
\end{align}
The largest $z$, denoted by $z_0$, which still allows for a real solution is the maximum of the right hand side for $Y$ from the interval $(-\min(\{k_i\}),0]$.
This interval follows from the discussion in the appendix of \cite{Plefka:02} about selecting the appropriate solution of Eq.~\eqref{eq:pastur}. The value of $Y$ at which $z_0$ is attained is denoted $Y_0$. By numerical optimization both $z_0$ and $Y_0$ can easily be found. The individual band edge $z_0$ improves on $x^2/4p$ by going beyond the quadratic approximation but it is still an infinite system result through the use of Pastur's theorem.

Hence we are now looking for a finite size correction to it.
The eigenvalue density $\rho$, when computed from the full equation \eqref{eq:pastur2}, still starts off with a square root singularity, i.e.
\begin{align*}
  \rho(\mu) \approx \gamma \sqrt{\mu-z_0}
\end{align*}
for $\mu$ close to $z_0$ and some constant $\gamma>0$. For a system of size $N$ the smallest eigenvalue $\mu_1$ will be roughly determined by the condition
\begin{align}
  N\int_{z_0}^{\mu_1}\rho(\mu)\,d\mu &= 1,
  \label{eq:mu1}
\end{align}
such that in our case
\begin{align*}
  \frac 23 N\gamma (\mu_1-z_0)^{3/2} &= 1,
\end{align*}
so $\mu_1 \approx z_0 + \left(\frac 23 N \gamma\right)^{-2/3}$. The second smallest eigenvalue $\mu_2$ can be calculated in the same way by replacing the right-hand side of Eq.~\eqref{eq:mu1} by 2, so $\mu_2 \approx z_0 + \left(\frac 13 N \gamma\right)^{-2/3}$.

The constant $\gamma$ can be calculated as follows. Tayor expansion to second order of the right hand side of Eq.~\eqref{eq:resmax} around the maximum gives
\begin{align*}
  z = z_0 - \frac 12 (Y-Y_0)^2 \frac{2\beta}{N} \sum_i (Y_0 + k_i)^{-3}
\end{align*}
such that
\begin{align*}
  Y &= Y_0 + i \sqrt{\frac{z-z_0}{\frac{\beta}{N} \sum_i (Y_0 + k_i)^{-3}}}
\end{align*}
for $z\ge z_0$. On the other hand $Y = \beta R(z) - z/\beta$ by definition. Comparison with Eq.~\eqref{eq:evdens} shows
\begin{align*}
  \gamma &= \frac{1}{\pi\beta^{3/2}}\left(\frac 1N \sum_i (Y_0 + k_i)^{-3}\right)^{-1/2}.
\end{align*}
This can be evaluated numerically since $Y_0$ is already known.

Thus we have calculated the individual band edge $z_0$ and its finite size correction for $\mu_1$,
\begin{align*}
  z_1 &= \beta\left(\frac{2}{3\pi} N\right)^{-2/3}\left(\frac 1N \sum_i (Y_0 + k_i)^{-3}\right)^{1/3},
\end{align*}
and for $\mu_2$,
\begin{align*}
  z_2 &= \beta\left(\frac{1}{3\pi} N\right)^{-2/3}\left(\frac 1N \sum_i (Y_0 + k_i)^{-3}\right)^{1/3}.
\end{align*}

\bibliography{refs}

\end{document}